# Probing Lyα Absorbers with Double Lines of Sight

Jane C. Charlton[1,3]

Christopher W. Churchill[2]

and

Suzanne M. Linder[1]

**Abstract**

Study of Lyα absorption lines in the spectra of double line of sight (DLOS) quasars holds the promise of diagnosing the nature of the structures that give rise to Lyα absorption. Based on simulations of DLOS with various separations through a single population of absorbers, four tests have been designed to diagnose absorber mass distributions (smoothly varying with radius or irregular), geometries (spherical or disk/slab–like), and kinematics (isotropic or systematic velocities). Applying the tests to existing data at $z \sim 2$ we find that: (1) The observed neutral hydrogen column density ($N_H$) distributions of Lyα lines coincident to both LOS are consistent with a smooth mass distribution. (2) Observed large anticoincident $N_H$ are not consistent with a single population of smooth spherical absorbers, which should exhibit sharp cut–offs at small $N_H$ in the $N_H$ distribution. (3) There is marginal evidence that the observed RMS value of velocity differences between coincident lines increases with DLOS separation as is expected for disk/slab–like absorbers that have systematic velocity fields. (4) The observed velocity dispersion along a single LOS is small compared to the RMS difference between widely separated LOS, which is not consistent with models of cloudlets moving isotropically within a spherical structure. Overall, only a smooth disk/slab–like model with systematic velocities remains consistent with the inferred properties of a single population of Lyα absorbers.

*Subject headings:* quasars: absorption lines — quasars: general

## Introduction and Motivation

Observations of coincident and anticoincident absorption lines[4] in the spectra of quasar pairs whose projected separations are small (hereafter double line of sight, DLOS) promise to place constraints on the nature of Lyα absorbers (Sargent, Young, & Schneider 1982, Foltz et al. 1984, Smette et al. 1992, Bechtold et al. 1994, Dinshaw et al. 1994 (hereafter F84, S92, B94, and D94)). We focus on observations of the neutral hydrogen column density $N_H$ and the velocity difference distributions of $z \sim 2$ Lyα forest lines along DLOS quasars. Our goal in this *Letter* is to demonstrate that various geometric and kinematic models give rise to discernibly different DLOS observations, which provide diagnostics of the mass distributions and kinematics of the structures. We do not attempt detailed predictions based upon rigorous models, but focus on idealistic, phenomenologically motivated models selected to illustrate expected behavior for more realistic scenarios. Monte–Carlo simulations of DLOS with various separations are presented for a spherical structure with an isotropic velocity field, a constant rotation "disk", and a flattened "slab" with constant velocity flowing radially toward or away from a central point (planar inflow/outflow). The resulting model $N_H$ distributions and velocity differences of the absorption lines are compared to current DLOS observations. We argue that the most consistent single population model is a flattened disk/slab–like structure with a $N_H$ distribution that smoothly falls off with radius and with some level of systematic motion in the plane.

The sizes of Lyα absorbers, inferred from the fraction of lines with coincident redshifts in both spectra, are too large to be explained by most traditional models. Assuming a spherical population of identical structures, B94 and D94 derived a most probable radius of $R \sim 80~h^{-1}$ kpc ($h = H_0/100$ km s$^{-1}$ Mpc$^{-1}$, $q_0 = 0.5$) for Lyα absorbers at $z \sim 2$ from their studies of Q1343+266 A,B. At redshifts $0.5 \leq z \leq 0.9$, coincident lines in the spectra of the more highly separated ($D \sim 340~h^{-1}$ kpc) Q0107–025 A,B provide an even larger strict lower limit of $D/2$ on the absorber radius (Dinshaw et al. 1995). These low redshift data should be considered separately since it is a matter of current debate whether high and low redshift Ly$\alpha$ absorbers comprise a single population. Since two of the four coincident lines in Q0107−025 A,B fall in the Lyman limit regime this letter will focus on the $z \sim 2$ population.

---

[4] Roughly, coincident lines are those which appear in both spectra of a close pair, whereas anticoincident lines are those which are observed in one of the spectra but have no detectable counterpart in the other spectrum.

The distributions of equivalent widths and velocity differences of coincident absorption lines at $z \sim 2$ show evidence of becoming less correlated with increasing DLOS separation. UM673 A,B, a lensed system with LOS separations ranging from $D \sim 0 - 2\ h^{-1}$ kpc, yields $\approx 50$ absorption lines in each spectrum. All but two[5] of the absorption lines are coincident along the two LOS, with equivalent widths and LOS velocities that are identical within errors (S92). Q2345+007 A,B has LOS separation $D \sim 11\ h^{-1}$ kpc, assuming a lensed system with $z_{\rm lens} = 1.49$ (Fischer et al. 1994). If we account for likely metal lines (Steidel & Sargent 1991), eleven lines are coincident and one is anticoincident (F84). The equivalent widths of the coincident lines are less correlated than those of UM673 A,B. The LOS RMS velocity difference is 25 km s$^{-1}$, but within the experimental errors this is consistent with 0 km s$^{-1}$ (F84). Q1343+266 A,B, a physical pair with separation $D \sim 40\ h^{-1}$ kpc, yields $\approx 10$ absorption lines in each LOS. Although the B94 and D94 line lists differ in detail (likely due to different spectral resolutions), significant variations in equivalent widths between the two LOS are observed (cf. Fig 3 of D94). D94 find an RMS velocity difference of $65 \pm 24$ km s$^{-1}$ among six coincident absorption lines.

In principle, coincident Ly$\alpha$ absorption lines could arise either because a continuous structure covers both LOS, or because a "cluster" of smaller absorbers intercepts both LOS. Consider four idealized models of Ly$\alpha$ absorbers: 1) a continuous mass distribution in a spherical structure, 2) discrete absorbers ("cloudlets") distributed in a spherical structure, 3) a continuous mass distribution in a disk/slab–like structure, and 4) discrete cloudlets distributed in a disk/slab–like structure. Based upon the large fraction of coincidences seen in Q1343+266 A,B (B94, D94), a cloudlet scenario requires near unit covering factor for column densities greater than detection thresholds. For the disk geometry, a covering factor near unity implies a filling factor near unity, so that little difference exists between the latter two models.

Continuous mass distributions likely result in a monotonically decreasing $N_H(r)$, where $r$ is the radius of the structure. At the opposite extreme, discrete cloudlets within a structure could give rise to an irregular $N_H$ distribution, independent of $r$. The kinematics within absorbing structures may range from highly systematic (e.g. rotation or planar inflow/outflow) to quite "random" (isotropic). To bracket a reasonable range we study two models with smoothly varying $N_H(r)$: (1) an isotropic velocity distribution in a spherical absorbing structure of radius $R$, which we interpret as discrete cloudlets moving isotropically with velocities $\sim (GM/R)^{1/2}$, where $M$ is the total mass of the absorbing structure, and (2) disk/slab–like structures of radius $R$ with either a constant rotation curve or a constant planar inflow/outflow with respect to a central point. We compare these models to a model with an irregular $N_H$ distribution. The smooth "isotropic sphere" serves to illustrate mini–halo class models (Rees 1986, Miralda-Escudé & Rees 1993), the "rotating disk" reflects the expected Ly$\alpha$ absorption properties of Vanishing Cheshire Cat (VCC) galaxies (Salpeter 1993, Salpeter 1995, Salpeter & Hoffman 1995), and the planar inflow/outflow "slabs" loosely illustrate Zeldovich pancakes (Cen et al. 1994, Shandarin et al. 1995, Hernquist, Weinberg, & Katz 1995).

### Monte–Carlo Simulations and Model Results

We assume all Ly$\alpha$ forest absorption lines at $z \sim 2$ are produced by a single population of absorbing structures. Thus, the observed column density distribution $f(N_H)dN_H \propto N_H^{-\beta}dN_H$ must be recovered from the $N_H(r)$ distribution across the absorbing structure, following $N_H(r) \propto r^{2/(1-\beta)}$. We use $\beta = 1.5$ (Petitjean et al. 1993, Rauch et al. 1993), which results in $N_H(r) \propto r^{-4}$. Using Monte–Carlo techniques, we simulate absorption properties along two LOS (A and B). We use the parameter $d = D/R$ to represent the ratio of the DLOS separation to the radius of the absorbing structure, and examine the values $d = 0.01, 0.1, 0.25, 0.5, 1.0$, and $1.5$. For each LOS, we compute a normalized column density $N = N_H(r)/N_H(R)$ and a normalized LOS velocity $V = V_{\rm los}/V_*$, where $N_H(R)$ is the neutral column density at the absorber "edge", $V_{\rm los}$ is a velocity vector projected along the LOS, and $V_*$ is the velocity vector magnitude. For comparison with observations, $N_H(R)$ represents the limiting column density of a sample.

---

[5] The two possible anticoincidences may be a MgII doublet.

Simulations for the disk geometry account for random orientations on the plane of the sky, which result in elliptical cross–sections. In this case, $N_H(r)$ is the column density at the radial distance *on* the disk/slab, given by $r = l \left\{ \cos^2 \alpha + (\sin \alpha / \cos i)^2 \right\}^{1/2}$, where $l$ is the LOS distance from the ellipse center, $i$ is the inclination subtended between the perpendicular axis of the disk and the plane of the sky, and $\alpha$ is the angle subtended between a vector pointing to the LOS and the semi–major axis of the ellipse. For rotational motion, the LOS velocity is given by $V_*(l/r)\cos \alpha \sin i$, where $V_*$ is the disk circular velocity. For planar inflow/outflow, $V_*$ represents the constant velocity magnitude and the LOS velocity is $\pm V_*(l/r) \sin \alpha \tan i$.

In Fig. 1, we present the distribution $\log(N_A)$ versus $\log(N_B)$, which we call $f(N_A, N_B)$. An irregular $N_H$ distribution (independent of $r$), in which $f(N_A, N_B) = N_A^{-1.5} N_B^{-1.5}$, is shown in panel (a). The smooth $N_H(r)$ distribution is illustrated in panels (b–f), where panel (d) shows a spherical absorbing structure for $d = 0.25$ and panels (b), (c), (e), and (f) show randomly oriented disks for $d = 0.01$, 0.1, 0.5, and 1.0. There is no expected correlation between $N_A$ and $N_B$ for the irregular $N_H$ distribution, which yields an $f(N_A, N_B)$ independent of $d$. The smooth spherical structure yields an increased probability density along the distribution "boundaries". The higher density arises from the large range of random angles that locate LOS B at similar radii, yielding similar $N_H$. Random inclinations of the smooth disks result in $(N_A, N_B)$ pairs that lie outside the regions of increased probability density. For small $d$, the coincident LOS are at virtually the same radii within a structure, so that the $N_H$ are highly correlated. As $d$ increases, the correlation weakens until at $d \geq 1$ the second LOS no longer passes through the structure at $N_B \geq N_H(R)$ if the first LOS has large $N_H$ (small $r$).

The $N_H$ distributions of anticoincidences $f(N_{ac})$, shown in Fig. 2, exhibit remarkable differences between the smooth spheres and smooth disks. For spherical structures with $d < 1$, the $f(N_{ac})$ sharply cut off at $N_{ac} = N_H(R)\,(1-d)^{-4}$. For disks, the $f(N_{ac})$ follow a power law for large $N_{ac}$, but exhibit increased probability densities at low columns that are more pronounced for smaller $d$. The latter arise due to increased probability for an anticoincidence in the case of intermediate to high inclination, when the impacting LOS is likely to have small $N_H$ (hit at large $r$ on the disk). At $d \geq 1$ the $f(N_{ac}) \propto N_{ac}^{-1.5}$, and are indistinguishable from the $N_{ac}$ distribution arising from irregular structures.

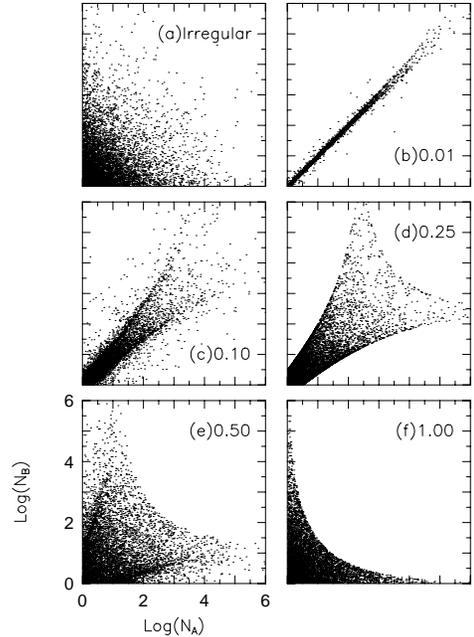

Fig. 1.— The distributions of column densities $f(N_A, N_B)$ from our models. — (a) an irregular mass distribution of "cloudlets" (no dependence on $d$), — (b) and (c) the $d = 0.01$ and 0.1 cases for a smooth disk/slab–like structure with column distribution $N_H(r) \propto r^{-4}$, — (d) the $d = 0.25$ case for the smooth spherical structure, — (e) and (f) the $d = 0.5$ and 1.0 cases for the smooth disk/slab–like structure. For the smooth mass distribution models, note the increased probability density "ridges" astride the $N_A = N_B$ correlation line, and that as $d$ increases, the $(N_A, N_B)$ pairs become less correlated with the maximum column density of a correlated pair given by $N_H(R)(d/2)^{-4}$.

In Fig. 3, we show the distribution of velocity differences $f(\Delta V = |V_A - V_B|)$ between LOS for the isotropic sphere ($d$ independent), the rotating disk, and the planar inflow/outflow slab. If a significant systematic component to the velocities exists in the disk/slab, an observable change in $f(\Delta V)$ will occur as a function of $d$, with $f(\Delta V)$ highly peaked around $\Delta V = 0$ for small $d$. The distribution shapes of the rotation and planar inflow/outflow cases are virtually indistinguishable for $d \lesssim 1$, except that the maximum velocity difference for the planar inflow/outflow case has an inclination dependent cutoff at $\Delta V_{\max} = 2V_\ast \left\{1 - (d/2)^2\right\}^{1/2}$. As $d$ increases over the range $1 \lesssim d \leq 2$, the decreasing $\Delta V_{\max}$ for the planar inflow/outflow case becomes clearly distinct from the rotation case with $\Delta V_{\max} = 2V_\ast$.

**Diagnostics from DLOS Observations**

As an example of an application of the tests in the previous section, we consider the current observational data at $z \sim 2$ (noting that these data are limited by small sample sizes, confusion due to blending, and uncertain $N_H$ on the flat part of the curve of growth). Our maximum likelihood analysis of the D94 data (following McGill (1990)) yields a most probable size of $R = 74\,h^{-1}$ kpc for spherical absorbers, and $R = 106\,h^{-1}$ kpc for disks. Adopting this value for the size of our single population of absorbers, we have $0 \leq d \leq 0.025$ over the observed redshift range ($2.1 \lesssim z \lesssim 2.7$) along UM673 A,B, and at $z \sim 2$, we have $d \sim 0.1$ for Q2345+007 A,B[6] and $d \sim 0.5$ for Q1343+266 A,B.

From coincident lines, the $f(N_A, N_B)$ distinguish between structures with smoothly varying monotonically decreasing $N_H(r) \propto r^{-4}$ and those with irregular $N_H$ distributions. Figure 1 (b–e) shows the predicted progression from strongly correlated $f(N_A, N_B)$ for UM673 A,B to relatively scattered for Q1343+266 A,B, which is generally consistent with the observations (F84, S92, B94, D94). The level of $f(N_A, N_B)$ correlation in the UM673 A,B data (Fig. 5a of S92) is not consistent with $N_H$ distributed irregularly on kpc scales. The Q2345+007 A,B data (Fig. 6 of S92) and the Q1343+266 A,B data (Fig. 3 of D94) are scattered about the correlation line, consistent with smoothly varying $N_H$ distributions on scales of tens of kpc within Lyα absorbing structures, and not with an irregular distribution. However, a higher level of correlation for large $d$ cases could arise if we relax the assumption of a single population such that each structure has an $N_H$ gradient smaller than $r^{-4}$. Scatter would be added to model predictions if we included a range of $d$ values resulting from variation in structure size and in DLOS separation with redshift.

---

[6] Fischer et al. (1994) and Steidel & Sargent (1991) find evidence that Q2345+007 A,B is a lensed pair. If it is *not* lensed, then we have $d \sim 0.5$, similar to Q1343+266 A,B.

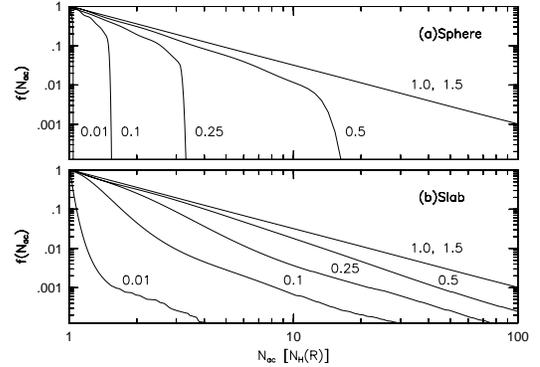

Fig. 2.— The distribution of anticoincident column densities $f(N_{ac})$. All distributions have been normalized at $N_H(R)$. Each curve is labeled by its corresponding $d$ parameter. — (a) the smooth spherical absorbing structures. — (b) the smooth disk/slab-like absorbing structures.

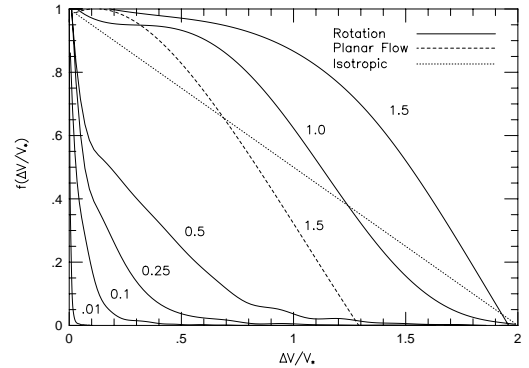

Fig. 3.— The distribution of LOS velocity differences $f(\Delta V = |V_A - V_B|)$. All distributions have been normalized at $\Delta V = 0$. Each curve is labeled by its corresponding $d$ parameter. The solid curves are for smooth disk/slab-like structures with constant rotational velocities. The dashed line is for disk/slabs with constant planar inflow/outflow velocities. The distribution arising from planar inflow/outflow exhibits a sharp cut off at $\Delta V/V_\ast = 2\left\{1 - (d/2)^2\right\}^{1/2}$, and becomes distinguishable from the rotating disk for $d > 1.0$. The dotted line is for smooth spherical structures with isotropic "cloudlet" velocities, which give $d$ independent distributions.

From anticoincident lines, the $f(N_{ac})$ distinguish between irregular, smooth spherical, and smooth disk/slab–like structures for $d < 1$. As seen in Fig. 2, smooth spheres yield a vanishingly small fraction of anticoincidences at large column densities, while for $d \lesssim 0.2$ smooth disks yield a significantly smaller fraction of anticoincidences than do irregular $N_H$ distributions. For example, if $d = 0.1$ we find $\sim 12\%$ of the anticoincident lines that arise from smooth disks have $N_H \geq 2N_H(R)$, whereas $\sim 71\%$ are above this $N_H$ in the case of irregular distributions. Significant differences between smooth spheres and disks persist for larger $d$ values. For $d = 0.5$, no anticoincidences with $N_{ac}$ greater than $\sim 15N_H(R)$ should be observed if the geometry is spherical as compared to 26% for disk models. Since, for $d = 0.5$, $\approx 44\%$ of the absorption lines are anticoincidences, $\approx 11\%$ of the total number of absorption lines should have $N_{ac} \geq 15N_H(R)$ in the case of a disk geometry. For small samples, DLOS separations in the range $0.25 \leq d \leq 0.5$ are likely to provide the greatest leverage.

In current studies, there are two anticoincident columns from which we infer that a single population of smooth spherical $N_H(r)$ Lyα absorbers is not consistent with the data. First, it is important to account for the fact that in an equivalent width limited sample the minimum detectable $N_H$ varies with wavelength (redshift). The minimum requirement for an anticoincidence is that a line of equal strength can be detected in the other LOS within a defined velocity/redshift window at the level chosen for the analysis. But even if the $N_H$ detection limit in a LOS is an order of magnitude below a detection in the other LOS, a biasing toward anticoincidences with large column densities may be occurring. For the $d = 0.1$ Q2345+007 A,B, F84 find an anticoincident line in spectrum A (line A1) with $N_H = 3 \times 10^{15}$ cm$^{-2}$ ($b = 30$ km s$^{-1}$). The B spectrum limiting column in this velocity/redshift window is $N_H = 6 \times 10^{14}$ cm$^{-2}$, a factor of 5 lower. With $D \sim 10$ kpc, a smooth sphere with $N_H(r) \propto r^{-4}$ should have a coincident line in spectrum B in the range $10^{15} \leq N_H \leq 10^{16}$ cm$^{-2}$. Here is a clear case in which an anticoincident line of $N_{ac}/N_H(R) \sim 5$ is seen when the $f(N_{ac})$ cut–off is $N_{ac}/N_H(R) \sim 1.5$ (see Fig. 2). For this anticoincident line to be consistent with a single population of spheres having $N_H(r) \propto r^{-\alpha}$, the column distribution would have to fall unreasonably steeply, with $\alpha \geq 15$. A similar argument holds for anticoincidence B1 from B94. Caution is in order, however, because an analysis of anticoincidences that relies on such small numbers is suspect to contamination of the Lyα sample by metal lines (Steidel & Sargent 1991).

From coincident lines, the $f(\Delta V)$ diagnose the degree of systematic motion in Lyα absorbers through analysis of correlations of the RMS velocity differences ($\Delta V_{rms}$) with $d$. Both the UM673 A,B ($d \sim 0.01$) and Q2345+007 A,B ($d \sim 0.1$) data yield $\Delta V_{rms}$ consistent with 0 km s$^{-1}$ (S92, F84). The observed $\Delta V_{rms}$ of 17 km s$^{-1}$ (Fig. 10 in S92) for UM673 A,B implies that $V_* \lesssim 10$ km s$^{-1}$, unrealistically quiescent for an isotropic sphere. However, a 10–20 km s$^{-1}$ spread is plausible for disk/slab–like structures in which the DLOS penetrate through a flattened geometry at some inclination. Systematic motion over large scales in the plane should result in a correlation of $\Delta V_{rms}$ with $d$, but due to the 25 km s$^{-1}$ error in the Q2345+007 A,B data, it is not possible to determine if the RMS velocity spread observed for Q2345+007 A,B is larger than that for UM673 A,B. However, the larger $\Delta V_{rms} = 65$ km s$^{-1}$ observed for Q1343+266 A,B (D94) does provide evidence for systematic motion. It should be possible to distinguish if the motion is predominantly rotational or planar inflow/outflow from DLOS with $d \gtrsim 1.5$, since the latter should exhibit a maximum cut off $\Delta V_{max} < 2V_*$.

Single LOS data can be compared to DLOS data to obtain further diagnostic information. Cowie *et al.* (1995) have resolved C IV $\lambda\lambda 1548, 1550$ lines associated with Lyα forest absorbers using HIRES (Vogt *et al.* 1994) on the Keck 10–m telescope, finding that the absorbers are comprised of a few cloudlets along the LOS. The cloudlet velocity dispersions $\sigma_{los} \approx 18$ km s$^{-1}$ along a single LOS are notably less than $\Delta V_{rms} = 65$ km s$^{-1}$, measured for the $D \sim 40$ $h^{-1}$ kpc pair Q1343+266 A,B (D94). The inequality $\Delta V_{rms} > \sigma_{los}$ is not consistent with a spherical absorbing struc-ture in which multiple cloudlets are moving isotropically ($\Delta V_{\rm rms}$ of the cloudlets between DLOS average out to be less than $\sigma_{\rm los}$ for the cloudlets along a single LOS). Many other models, such as spherically symmetric infall or an asymmetric velocity distribution for a spherical distribution of cloudlets, are not fully consistent with the data. The observed $\Delta V_{\rm rms} > \sigma_{\rm los}$ does, however, arise naturally in a model in which the motion is confined within the plane of a disk/slab–like structure.

The detailed distribution in position and velocity of the cloudlets responsible for Ly$\alpha$ absorption cannot be uniquely determined on the basis of current observations. However, some models are ruled out by our tests, while others remain consistent with the data. In particular, DLOS passing through cloudlets (blended in Ly$\alpha$) embedded in randomly inclined absorbing structures with flattened geometries and a smoothly declining $N_H(r)$ can give rise to the observed $f(N_A, N_B)$ and $f(N_{\rm ac})$ distributions, and to a LOS cloudlet velocity dispersion $\sigma_{\rm los} = 18$ km s$^{-1}$. If the kinematics is systematic as we have modeled, then the observed correlation (tentative) of $\Delta V_{\rm rms}$ with $d$ will hold true and the $f(\Delta V)$ will generally behave as shown in Fig. 3.

**Conclusions**

The new generation of 8–m class telescopes and efficient spectrographs will result in a more than ten-fold increase in the number of coincident and anticoincident lines, column densities will be directly measured, and blends will be separated. Expanded analyses similar to those discussed in §3 can then be pursued, comparing model predictions to the column distribution of pairs, $f(N_A, N_B)$ (Fig. 1), the distribution of anticoincident lines, $f(N_{\rm ac})$ (Fig. 2), and the distribution of velocity differences between pairs, $f(\Delta V)$ (Fig. 3). The discovery of more quasar pairs, and even groups, with separations ranging from $d = 0.01$ to $d = 1.5$, will allow the full range of diagnostics to be applied.

The simple models that we have considered here, though idealistic, should bracket the range of reasonable conditions comprising Ly$\alpha$ forest structures. These models provide clear predictions for correlations between the $d$ parameter and the distributions of coincident and anticoincident column densities and velocity differences. Based upon a loose interpretation of current data, a flattened structure with a component of systematic velocity (either rotation or planar inflow/outflow) is favored over a spherical structure with isotropic velocities. The large differences in the predicted distributions for the extreme models provides great optimism that the elusive answer to the questions of taxonomy, cosmogenesis, and general nature of Ly$\alpha$ forest *clouds* may soon be within our grasp.

This work was supported in part by NASA grant NAGW–3571 at Penn State. We thank N. Dinshaw, R. Guhathakurta, M. Keane, M. Rees, E. Salpeter, D. Schneider, C. Steidel, S. Vogt, and R. Wade for insightful suggestions, and acknowledge the hospitality of the Aspen Center for Physics, where many ideas were developed.

**REFERENCES**

Bechtold, J., Crotts, A.P.S., Duncan, R.C., and Fang, Y. 1994, ApJ, 437, L83 (B94)

Cen, R., Miralda–Escudé, J., Ostriker, J.P. & Rauch, M. 1994, ApJ, 437, L9

Cowie, L.L., Songaila, A., Kim, T–S, and Hu, E.M. 1995, AJ, 109, 1522

Dinshaw, N., Foltz, C.B., Impey, C.D., Weymann, R.J., and Morris, S. L. 1995, Nature, 373, 223

Dinshaw, N., Impey, C.D., Foltz, C.B., Weymann, R.J., and Chaffee, F. H. 1994, ApJ, 437, L87 (D94)

Fischer, P.F., Tyson, J.A., Bernstein, G.M., and Guhathakurta, P. 1994, ApJ, 431, L71

Foltz, C. B., Weymann, R. J., Röser, H.-J., and Chaffee, F. H. 1984, ApJ, 281, L1 (F84)

Hernquist, L., Weinberg, D., & Katz, N. 1995, preprint

McGill, C. 1990, MNRAS, 242, 544

Miralda–Escudé, J., Rees, M.J. 1993, MNRAS, 260, 617

Petitjean, P., Webb, J. K., Rauch, M., Carswell, R. F., and Lanzetta, K. 1993, MNRAS, 262, 499

Rauch, M., Carswell, R. F., Webb, J. K., and Weymann, R. J. 1993, MNRAS, 260, 589

Rees, M.J. 1986, MNRAS, 218, 25

Salpeter, E.E. 1993, AJ, 106, 1265

Salpeter, E. E. 1995, in *The Physics of the Interstellar Medium and Intergalactic Medium*, ed. A. Ferrara., C. Heiles, C. McKee, & P. Shapiro (PASP Conference Series), in press

Salpeter, E.E. and Hoffman, G.J. 1995, ApJ, 441, 51

Sargent, W.L.W., Young, P., and Schneider, D.P. 1982, ApJ, 256, 374

Shandarin, S. F., Melott, A. L., McDavitt, K., Pauls, J. L., and Tinker, J. 1995, Phys. Rev. Lett., 75, 7

Smette, A., Surdej, J., Shaver, P.A., Foltz, C.B., Chaffee, F.H., Weymann, R.J., Williams, R.E., and Magain, P. 1992, ApJ, 389, 39 (S92)

Steidel, C.C., and Sargent, W.L.W. 1991, AJ, 102, 1610

Vogt, S.S., *et al.* 1994, SPIE, 2198, 326